%% file: PeyresqHoevenaars_final.tex
\documentclass[12pt]{article}

\usepackage{amsmath,amssymb,amsthm}
\usepackage[pdftex]{graphicx,color}

\newtheorem{definition}{Definition}
\newtheorem{lemma}{Lemma}
\newtheorem{theorem}{Theorem}

\newcommand{\eq}{\begin{equation}}
\newcommand{\eeq}{\end{equation}}
\newcommand{\la}{\left\langle}
\newcommand{\ra}{\right\rangle}

\title{Frobenius manifolds and algebraic integrability}
\author{L.K. Hoevenaars\thanks{Research supported by VENI grant program of NWO}}
\date{}

\begin{document}
\maketitle

\abstract{We give a short review of Frobenius manifolds and algebraic integrability and study their intersection. The simplest case is the relation between the Frobenius manifold of simple singularities, which is almost dual to the integrable open Toda chain. New types of manifolds called extra special K\"ahler and special $F$-manifolds are introduced which capture the intersection.}

\section*{Introduction}
In the beginning of the '90s B. Dubrovin formalized the concept of a family of 2-dimensional topological field theories in terms of differential geometry. He showed that there is a 1-1 correspondence between such families and what he called Frobenius manifolds. The essential features of a Frobenius manifold are an associative commutative multiplication on each tangent space and a flat metric which is compatible with the multiplication. 
Furthermore there is a potentiality condition which allows one to express the structure constants of the multiplication in terms of a single function called the prepotential.
The associativity of the multiplication is equivalent to a set of third order nonlinear partial differential equations satisfied by the prepotential called the Witten-Dijkgraaf-Verlinde-Verlinde (or WDVV) equations. Dubrovin's achievements can be interpreted as constructing a geometrization of these equations in coordinate independent terms.

Frobenius manifolds have emerged in a wide range of contexts, e.g. singularity theory, quantum cohomology and complex structure moduli spaces of Calabi-Yau varieties. Manin, Hertling and others recognized that the concept of Frobenius manifold is too strict to capture a number of interesting structures in mathematics and physics and various relaxations have been introduced, the basic one being the concept of an $F$-manifold. By now a vast literature has formed and more information can be found for example in the books \cite{MANI:1999}, \cite{HERT-MARC:2004}. 

The present paper deals with the occurrence of a commutative associative multiplication in the context of algebraic integrability, which is not covered in the standard literature on Frobenius manifolds. 
The story begins with the seminal paper \cite{SEIB-WITT1:1994} by Seiberg and Witten in which the quantum version of $N=2$ supersymmetric Yang-Mills theory is solved. It had been known before \cite{WIT-LAUW-PROE:1985} that such theories can be solved by identifying the correct special K\"ahler manifold, which is what Seiberg and Witten did in a particular case. Subsequently it was realized in \cite{DONA-WITT:1996}, \cite{GORS-KRIC-MARS-MIRO-MORO:1995}, \cite{FREE:1999} that special K\"ahler manifolds and algebraically completely integrable systems are essentially the same things, which allowed for the solution of large classes of Yang-Mills theories by identifying the correct integrable system. 

Like Frobenius manifolds, any algebraically integrable system can be captured in terms of a single function again called the prepotential.
It was quickly realized by Marshakov, Mironov and Morozov \cite{MARS-MIRO-MORO:1996} that in some examples 
the prepotential satisfies the Witten-Dijkgraaf-Verlinde-Verlinde equations, thus making contact with the theory of Frobenius manifolds. This raises a number of questions which remain largely unsolved: for what class of integrable systems does this hold true? Can one find an appropriate coordinate free setting to cover this situation? Is there a more direct relation between 2-dimensional topological field theories and 4-dimensional Yang-Mills theories?

We will briefly comment on the current status of the first two of these questions. The study of explicit examples of algebraically integrable systems \cite{MARS-MIRO-MORO:1996, MARS-MIRO-MORO:1997, MARS-MIRO-MORO:2000, GRAG-MART:1999, HOEV-KERS-MART:2001, HOEV-MART2:2003} has learned that the prepotential does indeed satisfy the WDVV equations for the open Toda chain, the periodic Toda chain, relativistic Toda chain and various spin chains. In a future paper, we will show that the Neumann system is another example. On the other hand, it seems that it does {\it{not}} do so for systems of Calogero-Moser type \cite{MARS-MIRO-MORO:2000}. 
A reason for this, or any general pattern whatsoever, is still unclear at the moment. The only case where an explanation is given is the open Toda chain, which is related to the Frobenius manifold of a simple singularity through Dubrovin's concept of almost duality \cite{DUBR:2004}.
Clarity could be provided by finding an appropriate coordinate free characterization, i.e. a new type of manifold (close to Frobenius manifolds) that captures the WDVV equations in the context of algebraically integrable systems. Besides being a review, the present paper is an effort in that direction. 

The organization of the paper is as follows: the first section contains an introduction to Frobenius manifolds illustrated by examples associated with simple singularities. In section 2 we review the two related concepts of algebraic integrability and special K\"ahler manifolds. The relation is explained in section \ref{section special Kahler} and illustrated by the open Toda chain in section \ref{section Toda}. 
The final section contains a proposal for a new type of manifold, which we call a very special K\"ahler manifold, which we believe to be the appropriate one for studying associativity questions for integrable systems.
An even further specialization, which holds in the case of the open Toda chain, is called a special $F$-manifold and is an $F$-manifold with compatible flat metric.

We denote the tangent bundle of any complex manifold $B$ by $TB$, the holomorphic tangent bundle by ${\cal T}B$ and the projection onto the holomorphic part by $\pi^{(1,0)}$. The space of sections of a vector bundle $\pi:E\to B$ is denoted by $\Gamma(B,E)$ or simply by $\Gamma(E)$.

\section{Frobenius manifolds}
\label{section Frobenius}
Since the work of Dubrovin (see e.g. \cite{DUBR:1996}), it has become clear that any family of 2-dimensional Topological Field Theories is in one-to-one correspondence with a so-called Frobenius manifold. 
Roughly speaking, a Frobenius manifold is endowed with a flat metric and characterized by the presence of an associative commutative algebra structure on each tangent space which satisfies certain compatibility and integrability properties. 
This type of manifold has since appeared in many different contexts (e.g. singularity theory \cite{SAIT:1993, DUBR:1996}, quantum cohomology \cite{MANI:1999}, generalized deformations of complex structure \cite{BARA-KONT:1998} and integrable hierarchies \cite{KONT:1992, KRIC:1997, LEUR-MART:1999}) in many different guises. The literature on the subject is by now so vast that even the very definition of a Frobenius manifold has become somewhat unclear, since it is appropriate for the different contexts to drop some of the original requirements. 

The most basic notion is that of an $F$-manifold due to Hertling \cite{HERT:2002}
\begin{definition}
\label{Fman}
An $F$-manifold $(B,*,e)$ is an $n$-dimensional complex manifold $B$ together with a unital, associative, commutative multiplication $*:{\cal T}_bB \times {\cal T}_bB \rightarrow {\cal T}_bB$ for each holomorphic tangent space and a global unit vector field $e$. Moreover the multiplication satisfies
\eq
\label{Feq}
Lie_{X*Y}(*) = X*Lie_Y(*) + Y*Lie_X(*)
\eeq
\end{definition}
Here $Lie$ is interpreted as a derivation on tensor fields in the usual way. 
Explicit calculation shows that (\ref{Feq}) is equivalent to 
\begin{multline}
\label{id}
[X*Y,Z*W]-[X*Y,Z]*W-[X*Y,W]*Z-X*[Y,Z*W] 
-Y*[X,Z*W] + \\X*[Y,Z]*W + X*[Y,W]*Z  
+ Y*[X,Z]*W + Y*[X,W]*Z =0 
\end{multline}the left hand side of which is a 4-tensor. This fact allows one to check (\ref{id}) in local coordinates.

Endowing the manifold with a metric, one would like to have appropriate interrelations with the multiplication. One such relation is captured by the notion of a Frobenius algebra.

\begin{definition}
\label{Frobman}
A Frobenius algebra $(A,*,e,\la.,.\ra)$ is an associative commutative unital algebra $A$ with multiplication $*$ and unit $e$, together with a symmetric nondegenerate bilinear form $\la.,.\ra$ which is compatible with the multiplication:
\eq
\la X* Y,Z \ra = \la X, Y*Z \ra
\eeq
In other words, the 3-form $c(X,Y,Z)=\la X,Y*Z \ra$ is symmetric.
\end{definition}

\begin{definition}
A Frobenius manifold $(B,*,e,E,\la.,.\ra)$ is an $n$-dimensional $F$-manifold with a symmetric complex nondegenerate bilinear form $\la.,.\ra$ (henceforth called the metric) which makes each ${\cal T}_bB$ into a Frobenius algebra. Denoting the Levi-Civita connection for the metric by $\nabla$, one has moreover the following requirements
\begin{enumerate}
\item
The metric $\la.,.\ra$ is flat
\item
\label{coordinate}
The unit vector field $e$ is flat: $\nabla e=0$
\item
\label{potential}
$\nabla c$ is symmetric in all its (four) arguments 
\item
\label{Euler}
There is a vector field $E$ and a number $d$ such that

\begin{tabular}{l}
$Lie_E(*) =*$ \\
$Lie_E \la.,.\ra = d\la.,.\ra$
\end{tabular}
\end{enumerate}
\end{definition}
Here $c$ is as in definition \ref{Frobman} and $\nabla$ and $Lie$ act as derivations on tensors.
The vector field $E$ is called the Euler vector field.
Regardless of flatness of $e$, property \ref{potential} together with the Poincar\'e lemma implies that there exists locally a single holomorphic function $F$ such that 
\eq
c\left( X,Y,Z \right)=(XYZ)F
\eeq
for any triple $X,Y,Z$ of flat vector fields. Such a function $F$ is called a prepotential, characterizes the Frobenius manifold but is not unique: for example, any polynomial of degree 2 can be added to it.
Property \ref{coordinate} implies that one can choose the flat coordinates $\{t_k\}$ in such a way that $e=\frac{\partial}{\partial t_1}$ and therefore 
\begin{eqnarray}
c(\partial_{t_i} \partial_{t_j}\partial_{t_k}) &=& \frac{\partial^3 F(t_1,\dots t_n)}{\partial {t_i} \partial {t_j}\partial {t_k}}
\\
\la \partial_{t_i},\partial_{t_j} \ra &=& \la e*\partial_{t_i},\partial_{t_j}\ra = c(e,\partial_{t_i},\partial_{t_j})=\partial_{t_1}\partial_{t_i}\partial_{t_j}F
\end{eqnarray}
Naturally, not just any function can occur as a prepotential
\begin{lemma}
A holomorphic function $F(t_1,...,t_n)$ locally characterizes a Frobenius manifold if and only if it is quasihomogeneous and satisfies the following properties:
\begin{itemize}
\item
The matrix $[F_1]$ is constant and nondegenerate, where $[F_i]$ denotes the Hesse matrix of the first order derivative $\frac{\partial F}{\partial t_i}$.
\item
$F$ satisfies the Witten-Dijkgraaf-Verlinde-Verlinde differential equations (also called the associativity equations)
\eq
\label{WDVV}
[F_i] [F_1]^{-1} [F_j] = [F_j] [F_1]^{-1} [F_i] \qquad \forall i,j=1,...,n
\eeq
\end{itemize}
\end{lemma}
The quasihomogeneity takes care of the Euler vector field, the first property ensures flatness and nondegeneracy of the metric and a constant unity and the second is equivalent to associativity of the multiplication on the tangent space.

For future reference we also introduce the intersection form $(.,.)$, which is another flat metric naturally associated to any Frobenius manifold.
\begin{definition}
The multiplication $*$ is also defined on the cotangent spaces via duality. The intersection form $(.,.)$ is a bilinear form on the cotangent spaces given by
\eq
\label{intersection}
(\omega_1,\omega_2) = \iota_E \left( \omega_1 * \omega_2 \right)
\eeq
By abuse of notation, we denote also by $(.,.)$ the bilinear form on the tangent bundle.
\end{definition}
By a result of Dubrovin, $(.,.)+\lambda \la.,.\ra$ form a flat pencil of metrics: this bilinear form is flat for any parameter $\lambda$.

\subsection{An example: Saito's theory for simple singularities}
\label{section singularities}
Consider a simple singularity of ADE type, realized explicitly in terms of the zero level set of the polynomial $s(x_1,x_2,x_3)$ as in table \ref{table singularities}. It has unfoldings $S(x_1,x_2,x_3,b_1,...,b_n)$ where $n$ denotes the rank of the Lie algebra of ADE type, and $S(x_1,x_2,{x_3},0,...,0)=s(x_1,x_2,{x_3})$. 
Any unfolding defines a generalized Milnor fibration
\eq
\left\{ S(x_1,x_2,{x_3},b_1,...,b_n)=0 \right\} \longrightarrow B
\eeq
where $B=\mathbb{C}^n\backslash \Delta$ and $\Delta$ (the discriminant) is the set of points $b=(b_1,\dots, b_n)\in \mathbb{C}^n$ for which the fiber is singular.
An important invariant under certain changes of coordinates of a singularity is its Jacobi ring
\eq
\label{Jacobi}
J := \mathbb{C}[x_1,x_2,{x_3}] / ( \partial_{x_1} s, \partial_{x_2} s,\partial_{x_3} s )
\eeq
where the quotient is taken over the ideal spanned by the partial derivatives of $s$.
An unfolding is called semi-universal iff the Kodaira-Spencer type map
\begin{eqnarray}
\kappa: \Gamma({\cal T}B) &\to&  J \\
\kappa \left(\sum_k X_k\partial_{b_k}\right) &=& \sum_k X_k\frac{\partial S}{\partial b_k}
\end{eqnarray}
is an isomorphism of vector spaces (the unfoldings in table \ref{table singularities} are of this type). 
This isomorphism defines a natural multiplication on the holomorphic tangent spaces ${\cal T}_bB$ via pull-back.
For simple singularities, the multiplication is unital and one can arrange the deformation to be such that $e=\partial_{b_1}$ which maps under $\kappa$ to the equivalence class of $1$ in $J$. In other words, $\frac{\partial S}{\partial b_1}=1$. This defines the multiplication of a Frobenius manifold structure on $B$.
\begin{definition}
The metric (complex bilinear symmetric 2-form) $\la.,.\ra$ on $B$ is defined by
\eq
\label{metric}
\la X, Y\ra := \int_{\cap_{k=1}^3  \mid \partial_{x_k} S  \mid=\epsilon} \frac{\kappa(X) \kappa(Y)dx_1\wedge dx_2  \wedge d{x_3}}{\partial_{x_1} S \partial_{x_2} S\partial_{x_3} S}
\eeq
where $\epsilon$ is chosen small enough so that $\cap_{k=1}^3  \left| \partial_{x_k} S  \right|=\epsilon$ defines a real three-cycle in $\mathbb{C}^3$ around an isolated singularity of the integrand. 
\end{definition}
Since $S$ is a Morse function, $\cap_k \{ \partial_{x_k}S=0\}$ consists of isolated points $p_1,\dots p_n$ and the sum over all finite such points is understood. The multivariable residue theorem (see e.g. \cite{GRIF-HARR:1994},\cite{AIZE-YUZH:1983}) states that
the sum over all such points (including those at infinity) add up to zero, and that the sum over the finite points equals
\eq
\la X,Y\ra=\sum_{p_k}\frac{\kappa(X)(p_k)\kappa(Y)(p_k)}{D^2S(p_k)}
\eeq
where $D^2S$ denotes the Hessian determinant of $S$. This gives two alternative methods for calculating the metric: either using the finite points, or by calculating the residue at infinity.

It is easily seen that $\la.,.\ra$ is compatible with the product structure of (\ref{Jacobi}). 
In order to prove that the metric and multiplication give $B$ the structure of a Frobenius manifold, one still needs the following properties:
\begin{itemize}
\item
the metric is nondegenerate and flat
\item
$\nabla e=0$
\item
${\nabla}c$ is symmetric
\item
there is an Euler vector field
\end{itemize}
General proofs are contained for example in \cite{SAIT:1993}, \cite{DUBR:1998}. As a practical shortcut, we will give the formulae for the flat coordinates and for $c$. One can then check explicitly that in the flat coordinates, the metric becomes constant and nondegenerate. One also finds $\nabla e=\nabla \partial_{t_1}=0$. Since $c(\frac{\partial}{\partial t_i},\frac{\partial}{\partial t_j},\frac{\partial}{\partial t_k})$ is polynomial, it can be easily integrated which both shows that ${\nabla}c$ is symmetric and gives an explicit expression for the prepotential. 

\begin{table}[t]
\begin{tabular}{c|l|l}
$\mathfrak{g}$ & $s(x_1,x_2,x_3)$ & $S(x_1,x_2,x_3,b)$  \\
\hline
$A_n$ & $ x_1^{n+1} + x_2^2 + x_3^2$ & $ x_1^{n+1} + x_2^2 + x_3^2 + b_n x_1^{n-1} + b_{n-1}x_1^{n-2} + \dots +b_1$  \\
\hline
$D_n$ & $  x_2^{n-1} -x_1^2x_2 + x_3^2$ & $ x_2^{n-1}  - x_1^2x_2 + x_3^2 + b_n x_1 + b_{n-1}x_2^{n-2} + b_{n-2}x_2^{n-3}+ \dots +b_1$ 
 \\
\hline
$E_6$ & $ x_1^4 + x_2^3 + x_3^2$    & $ x_1^4 + x_2^3 + x_3^2 + b_6x_1^2x_2 + b_5x_1x_2 + b_4 x_1^2+ b_3 x_2+b_2x_1+b_1$ \\
\hline
$E_7$ & $ x_1^3x_2 + x_2^3 + x_3^2$ & $ x_1^3x_2 + x_2^3 + x_3^2 + b_7x_1x_2^3 + b_6 x_1x_2^2+b_5x_2^2+b_4x_1x_2+b_3x_2+b_2x_1+b_1$ \\
\hline
$E_8$ & $ x_1^5 + x_2^3 + x_3^2$      & $ x_1^5 + x_2^3 + x_3^2 + b_8x_1^3+x_2+b_7x_1^3+b_6x_1^2x_2+b_5x_1x_2+b_4x_1^2+b_3x_2+b_2x_1+b_1$ \\
\end{tabular}
\caption{Semi-universal unfoldings of the simple singularities.}
\label{table singularities}
\end{table}

The flat coordinates are given by
\eq
\label{flat}
t_k = b_k + P_k(b_{k+1},\dots, b_n)
\eeq
The polynomials $P_k$ can be found algorithmically as follows: assign degrees to the $b_k$ and to $x_1,x_2,x_3$ so as to make $S$ quasihomogeneous. Equation (\ref{flat}) must respect the grading and $P_k$ consists of only a finite number of possible monomials. Requiring the metric to be constant in the coordinates $t_k$ essentially fixes the coefficients of these monomials.

The formula for $c$ is
\eq
c(X,Y,Z) = \int_{\cap_{k=1}^3  \mid \partial_{x_k} S  \mid=\epsilon} \frac{XS\;  YS \; ZS \; dx_1\wedge dx_2 \wedge dx_3} {\partial_{x_1}S\partial_{x_2}S\partial_{x_3}S} 
\eeq
which in the case of a type $A_n$ singularity reduces to
\eq
\label{residue}
c(X,Y,Z) = \int_{\mid \partial_{x} {\tilde S}  \mid=\epsilon} \frac{X{\tilde S}\; Y{\tilde S}\; Z{\tilde S}\;dx}{\partial_{x}{\tilde S}}
\eeq
where
\eq
{\tilde S}(x,b_1,\dots,b_n) = S(x_1,0,0,b_1.\dots ,b_n)
\eeq
Explicit expressions for flat coordinates and prepotentials can be found in the literature \cite{DIJK-VERL-VERL:1991, KLEM-THEI-SCHM:1992}.

The Euler vector field is given by
\eq
E = \sum_{k=1}^n d_{n+1-k} t_k \partial_{t_k}
\eeq
where $d_k$ are the degrees of the Lie algebra of the $ADE$ singularity in increasing magnitude. This Euler vector field assigns degree $n+1-k$ to $t_k$ and the prepotential is a quasihomogeneous polynomial with degree equal to the Coxeter number.

\section{Algebraic integrability and special K\"ahler geometry}
\label{section algebraic integrability}
It was observed by Marshakov, Mironov and Morozov \cite{MARS-MIRO-MORO:2000} that a structure similar to that of a Frobenius manifold occurs on the base space of some algebraically integrable systems. For any algebraically integrable system one can define a holomorphic function characterizing it and for some systems this function satisfies the Witten-Dijkgraaf-Verlinde-Verlinde equations (\ref{WDVV}). We will make this connection more precise in the next section, but first we review the twin notions of algebraic integrability and special K\"ahler geometry.

\subsection{Algebraic integrability}
In general, dynamical systems give rise to ordinary differential equations which are too difficult to solve. Integrability is a property of dynamical systems which ensures that they are solvable, in the sense that one can define so called action-angle variables by performing quadratures. The action variables are conserved quantities and the angle variables move linearly with time. If they move on a compact manifold a theorem by Liouville, improved by Arnol'd, sais that this manifold is a torus (the Liouville torus). This torus depends on the values of the time-independent action variables, which gives rise to a family of tori.
Passing to complex dynamical systems (as opposed to real ones), integrable systems often turn out to be algebraically integrable, meaning that the Liouville tori are complex abelian varieties. We now recall the abstract notion of an algebraically integrable system.

\begin{definition}
Consider a $2n$ complex dimensional symplectic manifold $A$ with symplectic form $\eta \in H^{2,0}(A,\mathbb{C})$ and an $n$ complex dimensional manifold $B$.
An algebraically integrable system is a proper holomorphic map $\pi:A \to B$ such that
\begin{enumerate}
\item
$\pi$ is lagrangian with respect to $\eta$.
\item
The fibers are algebraic, i.e. $\forall b\in B$ the fiber $A_b$ is an abelian variety.
\end{enumerate}
\end{definition}
As recalled for example in \cite{FREE:1999}, one can replace locally\footnote{The global obstructions were considered in \cite{DUIS:1980}, but we will not consider them in this paper.} the fibration $\pi$ with the fibration $T^*B\to B$ through a symplectomorphism. So the local data of the integrable system can be captured purely in terms of geometric structures on the base $B$ of the fibration. This structure on $B$ is called special K\"ahler geometry and is the topic of section \ref{section special Kahler}.

\subsection{Lax pairs and spectral curves}
It often happens that the fibration of abelian varieties appearing in the definition of an algebraically integrable system is given by the (Prym subvarieties of) Jacobians of a family of algebraic curves. Such curves are called spectral curves and are usually given explicitly in terms of a {\it{Lax pair with spectral parameter}}. 
Before turning to the specific example we have in mind, we briefly review this common construction of algebraically integrable systems. 

We will take a dynamical system to be a $2n$ complex dimensional symplectic manifold $(A,\eta)$ together with a function $H\in C^{\infty}(A)$ called the Hamiltonian. Associated to the Hamiltonian is a vector field $X_H$ implicitly defined by
\eq
\iota_{X_H}\eta=dH
\eeq
In terms of Darboux coordinates $(p_i,q_j)$ the symplectic form is expressed as
\eq
\eta = \sum_{k=1}^n dp_k \wedge dq_k
\eeq
The equations of motion associated with the time flow of the one-parameter group generated by $X_H$ are given by Hamilton's equations
\begin{eqnarray}
\frac{dp_k}{dt} &=& X_H(p_k) = \eta(X_H,X_{p_k}) \\
\frac{dq_k}{dt} &=& X_H(q_k) = \eta(X_H,X_{q_k})
\end{eqnarray}
Suppose the equations of motion are equivalent to a set of matrix equations
\eq
\frac{dL}{dt} = [P,L]
\eeq
where $P,L$ are matrices whose entries depend on the coordinates and momenta of the system (and hence on time). Then it is easily seen that all the $Tr(L^k)$ and therefore the spectrum of $L$ are time-independent. The maximum number of time-independent elements $b_i\in C^{\infty}(A)$ which are in involution, i.e. $\eta(X_{b_i},X_{b_j})=0$, is $n$. In fortunate cases one can find this many independent ones among the various $Tr(L^k)$ which ensures that one is dealing with an integrable system, the $b_i$ playing the role of action variables.

The matrices $P,L$ in the above construction are called a Lax pair, and $L$ the Lax matrix. If it depends on an auxiliary\footnote{One need not restrict attention to just a complex parameter. Interesting examples occur with $\zeta$ a coordinate on an algebraic curve \cite{HITC:1987}.} parameter $\zeta \in \mathbb{C}$ one can use the Lax matrix to construct a family of spectral curves
\eq
{\cal C}_b = \left\{ \det \left[L(\zeta,p,q)-x\cdot \mbox{id} \right]=0 \right\}
\eeq
The coefficients in $x,\zeta$ of this equation are time-independent functions $f_i(p,q)$ which can all be expressed in terms of the $b_i$.
This defines a fibration of curves ${\tilde \pi}: {\cal C}\to B$ where $B$ parametrizes the time-independent functions $b_i$ but omits those for which the corresponding fiber ${\cal C}_b$ is more singular than the generic fiber. 
It gets the structure of $\mathbb{C}^n \backslash \Delta$, where $\Delta$ is a discriminant locus. 
To see the definition of an algebraically integrable system in this construction, one notes that often the dynamics can be seen to take place on the Jacobians of the curves, and $\pi:\mbox{Jac}({\cal C})\to B$ is the more relevant fibration.
If the symplectic form on $\mbox{Jac}({\cal C})$ is given by an exact holomorphic 2-form $\eta=d\lambda$ the integrable system is called of Seiberg-Witten type. Pulling back $\lambda$ via the Abel-Jacobi map one obtains a meromorphic 1-form on the family $C$ whose restriction to the fibers is closed, and whose derivatives with respect to the base variables (defined using the Gauss-Manin connection) when restricted to the fibers are holomorphic. Such a 1-form is called a Seiberg-Witten form. 

This general picture, in which the Liouville tori are given by Jacobians of spectral curves, must frequently be adapted when one looks at specific integrable systems. One of the more common situations is the one in which the Jacobian is too big, and one needs to canonically identify a suitable abelian (Prym) subvariety of the Jacobian of the right dimension. There is a well-developed literature on this subject, see e.g. \cite{ADLE-MOER1:1980, MCDA:1988, DONA:1993}. On the other hand, it can happen that the Jacobian is too small and the spectral curve singular. In that case it can be more suitable to pass to the generalized Jacobian, see e.g. \cite{AUDI:2002} for a description for the geodesic flow on $SO(4)$. The open Toda chain, which is the subject of section \ref{section Toda}, has a similar description.

\subsection{Special K\"ahler manifolds}
\label{section special Kahler}
We introduce special K\"ahler manifolds in a coordinate free setting, following Freed's paper \cite{FREE:1999}.
\begin{definition}
\label{special Kahler}
A special K\"ahler manifold is a K\"ahler manifold $(B,\omega,I)$ with almost complex structure $I$ and K\"ahler form 
$\omega$ together with a flat torsion-free connection $\nabla$ such that
\begin{itemize}
\item
$\nabla \omega=0$
\item
$d_\nabla I=0$
\end{itemize}
\end{definition}
{\bf{Note: }}
Since not $\nabla I=0$ but rather $d_\nabla I=0$, the connection $\nabla$ is {\emph{not}} holomorphic, in the sense that $\nabla_X Y$ is in general not a holomorphic vector field if $X,Y$ are. The commutator $[X,Y]$ however is holomorphic:
\begin{lemma}
\eq
\label{lempi}
\pi^{(1,0)}[X,Y]=[X,Y] \qquad \qquad \forall X,Y \in \Gamma({\cal T}B)
\eeq
\end{lemma}
\begin{proof}
The torsion-freeness of the connection can be restated as
\eq
d_\nabla id=0
\eeq
where $id\in \Gamma(T^*B\otimes TB)$ is the identity mapping. Since $I=\frac{id-\sqrt{-1}\pi^{(1,0)}}{2}$ one finds
\eq
d_\nabla \pi^{(1,0)}=0
\eeq 
whence (\ref{lempi}) follows.
\end{proof}
Following Freed, we introduce flat real Darboux\footnote{w.r.t. the symplectic form $\omega$} coordinates $\{p_i,q_j\}$ and special holomorphic coordinates $z_i$ such that
\eq
\frac{\partial}{\partial z_i} = \frac{1}{2} \left( \frac{\partial}{\partial p_i} -\sum_j T_{ij} \frac{\partial }{\partial q_j} \right)
\eeq
One then shows from the definitions that $T_{ij}$ is the Hessian of a holomorphic function 
\eq
T_{ij} = \frac{\partial^2 {\cal F}}{\partial z_i \partial z_j}
\eeq
and ${\cal F}(z)$ is called a prepotential\footnote{Note that we have now introduced two seemingly completely different notions of prepotential -- one for Frobenius manifolds and one for special K\"ahler manifolds. The origin of the name lies in physics, where both  functions play similar roles.}.
The prepotential itself is not a canonical object. Its definition depends on various choices, such as the choice of Darboux coordinates which is subject to the action of an affine symplectic group acting on the vector $({z_1},...,{z_n}, {\cal F}_1,...,{\cal F}_{n})$, consisting of the holomorphic coordinates and the first order derivatives of ${\cal F}$.
The changes of ${\cal F}$ under such transformations can be quite dramatic.

Besides the K\"ahler form $\omega$ and the connection $\nabla$, the canonical object associated with a special K\"ahler manifold is a holomorphic symmetric 3-tensor $\Xi \in Sym^3({\cal T}B)$.
\begin{definition}
There is a canonically defined global section $\Xi \in Sym^3({\cal T}B)$ on any special K\"ahler manifold, given by
\eq
\Xi(X,Y,Z) = \omega(X,(\nabla \pi^{(1,0)})(Y, Z))
\eeq
\end{definition}
Since the K\"ahler form is of type $(1,1)$ and $X$ is holomorphic, the second argument $(\nabla \pi^{(1,0)})(Y, Z)=\nabla_YZ-\pi^{(1,0)}\left(\nabla_Y Z \right)$ should be anti-holomorphic which it is.
In terms of the local coordinates, we have the following expressions
\begin{eqnarray}
\label{omega}
\omega &=& \sum_{i,j} \mbox{Im} \left( \frac{\partial^2 {\cal F}}{\partial z_i \partial z_j} \right) dz_i \wedge d{\bar z}_j 
\\
\Xi    &=& \sum_{i,j,k} \frac{\partial^3 {\cal F}}{\partial z_i \partial z_j \partial z_k}dz_i \otimes dz_j \otimes dz_k
\label{jacobians}
\end{eqnarray}
and so it is the second and third order derivatives of the prepotential which carry the intrinsic information rather than the prepotential itself.

Now that both algebraic integrability and special K\"ahler geometry are introduced, we quote the theorem relating the two.
\begin{theorem}\cite{DONA-WITT:1996, FREE:1999}
Locally, the following data are equivalent
\begin{itemize}
\item
An algebraically integrable system $(\pi:A\to B,\eta)$
\item
A special K\"ahler manifold $(B,\omega,I,\nabla)$ together with a flat, Lagrangian lattice $\Lambda \subset T^*B$.
\end{itemize}
\end{theorem}
In the case of Seiberg-Witten type integrable systems the lattice is given by the period integrals of the Seiberg-Witten 1-form around the cycles of the spectral curves. The periods around $A$-cycles give the special K\"ahler variables and the $B$-cycles provide the prepotential
\begin{eqnarray}
z_i &=& \oint_{A_i}\lambda
\\
{\cal F}_j (z) &=& \oint_{B_j}\lambda
\end{eqnarray}

\subsection{Example: the $A_n$ open Toda chain}
\label{section Toda}
The description of the open Toda chain in terms of an algebraically integrable system using spectral curves is not entirely straightforward. The naive spectral curve obtained from a Lax pair with spectral parameter is rational, but it is not quite the correct spectral curve.
Probably the most practical way to go about it is to view the open chain as a limit of the periodic (or closed) Toda chain \cite{BRAD-MARS:2001}, which allows one to see that the spectral curve is really a singular curve obtained by glueing together two rational curves in a number of points.

\begin{figure}[htbp]
$\begin{tabular}{l@{\hspace{2cm}}r}
\resizebox{8cm}{!}{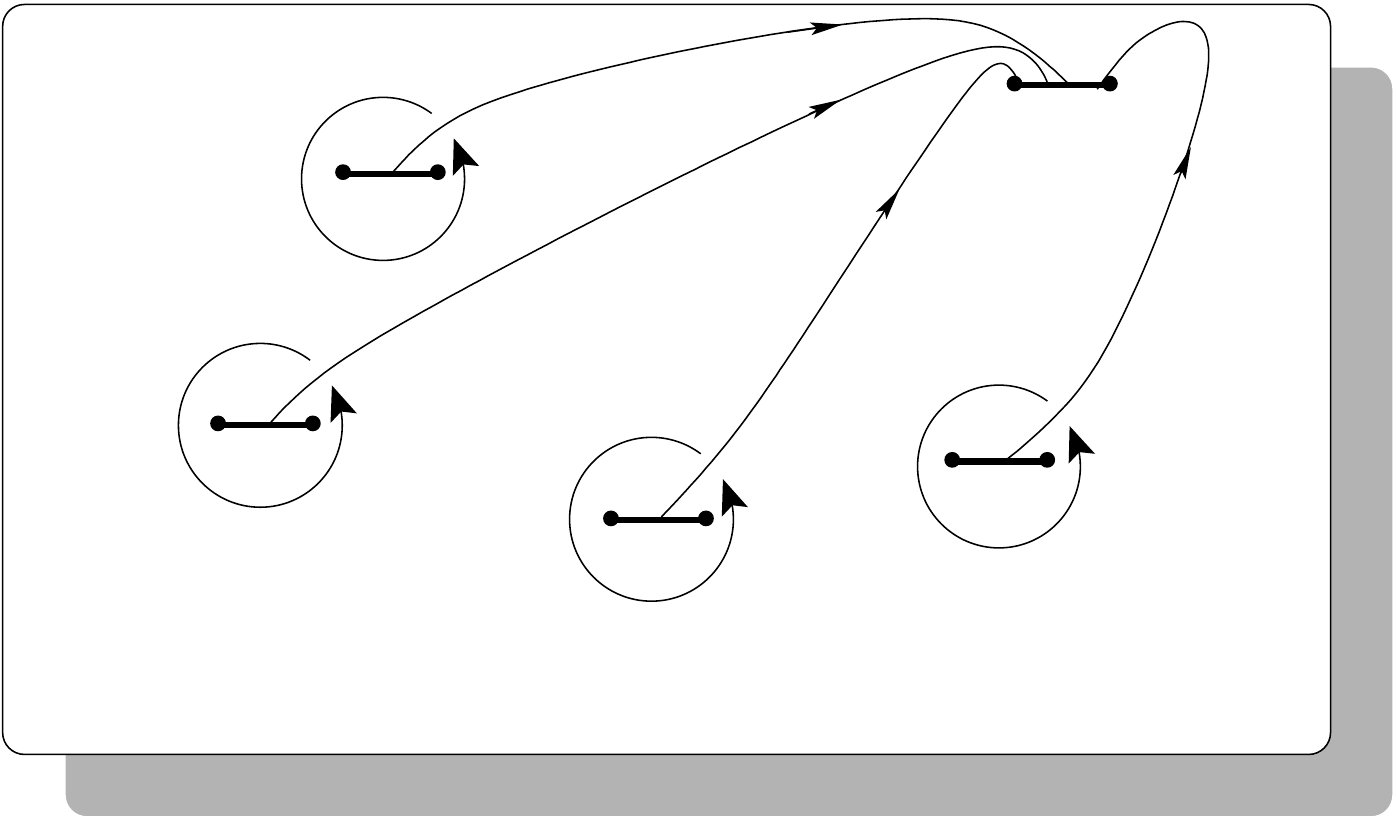}
&
\resizebox{8cm}{!}{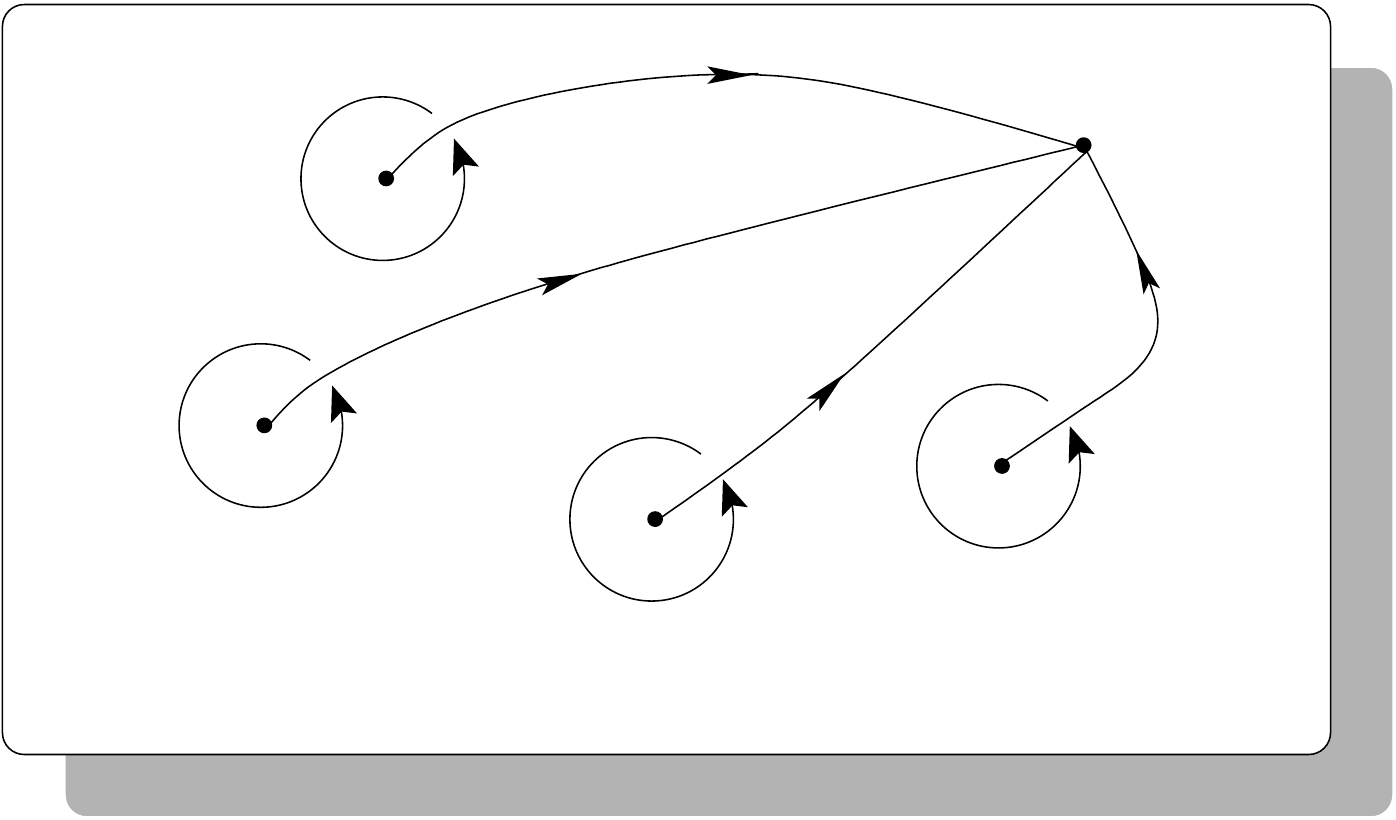}
\\
\resizebox{8cm}{!}{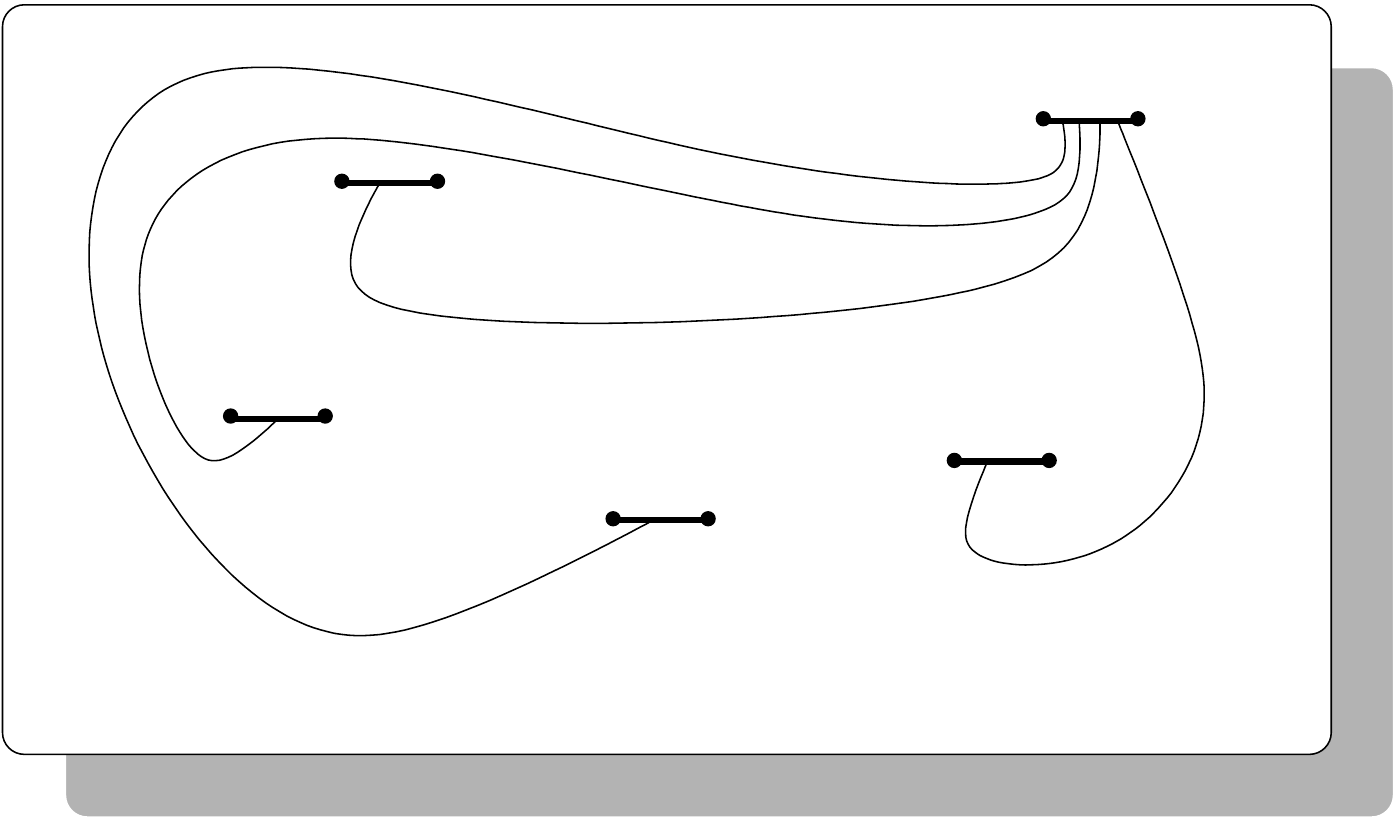} 
&
\end{tabular}$
\caption{On the left hand side the hyperelliptic spectral curve for the $A_4$ periodic Toda chain with canonical cycles. There are two sheets corresponding to the two solutions $\zeta_{\pm}(x)$ of (\ref{periodic Toda}). On the right hand side the corresponding open Toda chain rational curve with cycles}
\label{fig cycles}
\end{figure}

One starts with a semi-simple Lie algebra $\mathfrak{g}$, its root system $R$ and an irreducible representation $\rho$. The Lax pair $P,L$ for the open Toda chain is given by
\begin{eqnarray}
L &=& \rho \left( \sum_{i=1}^{\mbox{rank} (\mathfrak{g})} p_ih_i + e^{(\alpha_i,q)}e_{\alpha_i} + f_{\alpha_i} + \zeta e_{\alpha_0} \right)
\\
P &=& \rho \left(  \sum_{i=1}^{\mbox{rank}( \mathfrak{g})}f_{\alpha_i} \right)
\end{eqnarray}
Here the $\alpha_i$ are a set of simple roots, $\alpha_0$ is the highest root and $\{h_i,e_{\alpha_i},f_{\alpha_i}\}$ is a corresponding Chevalley basis for $\mathfrak{g}$.
For the Lie algebra of type $A_n$ and the fundamental representation, the spectral curve is of a particularly simple type
\eq
\label{type A}
\left\{ \zeta = {\tilde S}(x,b_1(p,q),...,b_n(p,q))=\prod_{k=1}^{n+1} (x-x_k(b)) \right\}
\eeq
where ${\tilde S}$ coincides with the corresponding semi-universal deformation of the singularity of type $A_n$ as in table \ref{table singularities} and therefore $\sum_k x_k=0$.
The correct way of looking at (\ref{type A}) however is to take two copies of this affine rational curve and identify them in the $n+1$ points $c_k=(x_k,\zeta_k)\in {\cal C}_{p,q}$ where ${\tilde S}(x_k,b)=0$.
The reason for this is that the $A_n$ {\it{periodic}} Toda chain has spectral curve
\eq
\label{periodic Toda}
\left\{ \zeta + \frac{\mu}{\zeta}= {\tilde S}(x,b_1(p,q),...,b_n(p,q)) \right\}
\eeq 
which is a double cover of the $x$-plane. The open Toda chain is obtained in the limit $\mu \to 0$, where the pairs of branch points given by solutions of
\eq
{\tilde S}(x)=\pm 2\sqrt{\mu}
\eeq
come together to form singular points of the curve
\eq
{\cal C}_b = \lim_{\mu\to 0} \left\{ \zeta + \frac{\mu}{\zeta}= {\tilde S}(x,b_1(p,q),...,b_n(p,q)) \right\}
\eeq
The homology of the curve is changed accordingly: the cycles of type $\delta$ survive to span the homology of ${\cal C}_b$ and the cycles of type $\gamma$ (the vanishing cycles) become trivial homologically (see figure \ref{fig cycles}).

The Seiberg-Witten differential $\lambda$ is given by
\eq
\lambda = x\frac{d\zeta}{\zeta}= \frac{x{\tilde S}_xdx}{{\tilde S}}
\eeq
which has singularities in the points\footnote{Of course upon compactification in $\mathbb{P}^2$ one finds another singularity at $[x,y,\zeta]=[0,0,1]\in {\mathbb{P}}^2$ which represents a point of the curve at infinity. This pole is of order two and has residue $\sum_k x_k=0$.} $c_1,...,c_{n+1}$. So although the $\gamma$ cycles are trivial homologically, the integral of the Seiberg-Witten differential around them is nonzero.
To give an idea of the special K\"ahler geometry thus constructed on $B$, we give an expression in local coordinates for the K\"ahler form $\omega$. We choose $\gamma, \delta$ cycles with canonical intersection numbers as in figure \ref{fig cycles}. Then we introduce
\eq
z_i = \oint_{\gamma_i}\lambda = x_i
\eeq
as the special coordinates on $B$. Next, we introduce
\eq
\label{B integrals}
{\cal F}_j = \oint_{\delta_j}\lambda
\eeq
There is a subtlety in evaluating these integrals. On the spectral curves of the periodic Toda chain ($\mu \neq 0$) the Seiberg-Witten 1-form $\lambda$ has no residues but on the rational curve this is no longer true. 
Whereas it is allowed to switch between the two representatives $-\log({\tilde S})dx$ and $\frac{x{\tilde S}_xdx}{S}$ of the cohomology class of $\lambda$ for the {\it{periodic}} Toda chain, the two give different periods when integrated on the rational curve of the {\it{open}} chain.
The proper way to interpret (\ref{B integrals}) is to evaluate the integrals over cycles on the periodic Toda curve and to take the limit \cite{HOKE-PHON:2002}. There is a short-cut however, which uses $\frac{x{\tilde S}_xdx}{S}$ for the $\gamma$-cycles as above and $-\log({\tilde S})dx$ to evaluate the $\delta$-periods directly on the rational curve.
\begin{lemma}\cite{MARS-MIRO-MORO:2000}
Integrating $-\log({\tilde S})dx$ over the non-compact $\delta$-cycles defines (\ref{B integrals}). The mixed derivatives ${\cal F}_{ij}=\frac{\partial{\cal F}_i}{dz_j}$ are symmetric up to a constant, so there exists a holomorphic function ${\cal F}(z)$ such that $\partial_{z_k}{\cal F}_{ij}$ are its third order derivatives and define a special K\"ahler structure on $B$.
\end{lemma}
Our choice of cycles leads to the prepotential
\eq
\label{Toda prepotential}
{\cal F}(z)= \frac{1}{2} \sum_{1\leq i<j\leq n} (z_i-z_j)^2 \log \left(e^{-\frac{3}{2}}(z_i-z_j)\right) + \frac{1}{2}\sum_{1\leq k \leq n}z_k^2\log(z_k) 
\eeq
We stress once more that the prepotential itself is {\it{not}} canonical. It depends on the choice of cycles and is therefore subject to transformations under the symplectic group. The K\"ahler form (\ref{omega}) however {\it{is}} canonical, in the sense that it is invariant.
A choice of cycles $\gamma_i,\delta_i$ is suitable as long as the corresponding local expression for $\omega$ leads to a positive metric. If not, a symplectic change of cycles is in order.

To complete the special K\"ahler geometry of the open Toda chain we mention that the connection $\nabla$ is given by the Gauss-Manin connection of the family ${\cal C}\to B$, which acts as a covariant derivative on the relative cohomology bundles.
An interesting feature of the prepotential of the open Toda chain of type $A_n$, whose desription is so close to the simple singularity of type $A_n$, is that a residue formula similar to (\ref{residue}) exists. 
As shown in \cite{MARS-MIRO-MORO:2000}, for any flat $X,Y,Z\in \Gamma({\cal T}_B)$
\eq
\label{residue Toda}
XYZ{\cal F} = \sum_{x=\infty} \mbox{res} \left[ \frac{\nabla_{X}\lambda\otimes \nabla_{Y}\lambda \otimes \nabla_{Z}\lambda}{dx\otimes \frac{dz}{z}}  \right] = \sum_{\partial_x{\tilde S}(x)=0} \mbox{res} \left[ \frac{X\log {\tilde S}\; Y\log{\tilde S}\; Z\log {\tilde S}\;dx}{\partial_x\log{\tilde S}}  \right]
\eeq
where the residues are taken on the rational curve (\ref{type A}).
The difference with this residue formula and the one given in (\ref{residue}) is that the coordinates on $B$ are chosen differently and that there is an extra factor ${\tilde S}^2$ in the denominator in (\ref{residue Toda}).

\section{Extra special K\"ahler manifolds}
We have defined a Frobenius manifold for the simple singularity of type $A_n$ and a special K\"ahler manifold for the open Toda chain of type $A_n$, both from the same basic manifold $B=\mathbb{C}^n\backslash \Delta$. These two structures are similar in the sense that they are both expressed locally through holomorphic prepotentials which satisfy residue formulae (\ref{residue}) and (\ref{residue Toda}). In this section we review the explanation of these similarities as given by Dubrovin from the Frobenius manifold point of view, and we introduce what we call extra special K\"ahler manifolds to address this issue from the special K\"ahler point of view. We will identify the position of extra special K\"ahler manifolds in the range between $F$-manifolds (little structure) and Frobenius manifolds (a lot of structure).

\subsection{Dubrovin's almost duality}
On a Frobenius manifold two flat metrics are naturally defined: the metric $\la.,.\ra$ and the intersection form $(.,.)$ of section \ref{section Frobenius}. Their Levi-Civita connections are part of a flat pencil of connections and indeed the two metrics play very similar roles on the Frobenius manifold. Indeed, one might ask whether a Frobenius manifold structure can be defined using $(.,.)$. Dubrovin \cite{DUBR:2004} showed that this is almost the case.

Given any commutative unital algebra $(A,*_e,e)$ with multiplication $*_e$ and unit $e$, together with an invertible element $E\in A$, one defines a new algebra $(A,*_E,E)$ by 
\eq
a*_Eb = (a*_e b)*_e E^{-1}
\eeq
This algebra is called a rescaling of the original algebra.
$E$ is the unit of the new algebra, and $*_E$ is associative if and only if $*_e$ is.

\begin{definition}
An almost Frobenius manifold $(B,*_E,E,e,(.,.))$ of degree $d\neq 1$ is a manifold with a Frobenius algebra structure $(T_bB,*_E,E,(.,.)))$ on each tangent space and
\begin{itemize}
\item
The metric $(.,.)$ is flat
\item
Defining $C(X,Y,Z)=(X*_EY,Z)$ one requires that
$\nabla C$ is symmetric in all its (four) arguments 
\end{itemize}
In terms of flat coordinates $z_k$ there exists a prepotential ${\cal F}(z)$ which satisfies the homogeneity condition
\eq
\sum_k z_k \frac{\partial {\cal F}}{\partial z_k} = 2{\cal F} + \frac{1}{1-d}(z,z)
\eeq
Moreover, there exists a special vector field $e$ such that ${\tilde z}\rightarrow e {\tilde z}$ acts by shifts $\nu \to \nu-1$ on the solutions of
\eq
\partial_{z_i} \partial_{z_j} {\tilde z} = \nu \sum_k C_{ij}^k \partial_{z_k} {\tilde z}
\eeq
where $C_{ij}^k$ are the structure constants of the Frobenius algebra.
\end{definition}

\begin{theorem}[DUBR:2004]
Given a Frobenius manifold $(B,*_e,e,E,\la.,.\ra)$ the intersection form is compatible with the rescaled multiplication $*_E$. The manifold $(B,*_E,E,e,(.,.))$ defines an almost Frobenius manifold structure. In the case of a simple singularity, the flat coordinates $z_k$ coincide with the special coordinates of the open Toda chain and the dual prepotential $\cal F$ coincides with the special K\"ahler prepotential. 
\end{theorem}

\subsection{Extra special K\"ahler manifolds and special $F$-manifolds}
For the Frobenius manifolds defined on the unfolding space of simple singularities, the almost dual manifolds coincide with the special K\"ahler structure of the open Toda chain. In particular, this means that the prepotential (\ref{Toda prepotential}) satisfies a version of the WDVV equations (\ref{WDVV}): if we denote by ${\cal F}_E$ the matrix of third order derivatives of ${\cal F}$ given by
\eq
\left[ {\cal F}_E \right]_{ij} = \sum_k z_k \frac{\partial^3{\cal F}}{\partial z_i\partial z_j \partial z_k}
\eeq
then the WDVV equations become
\eq
\label{gen WDVV}
{\cal F}_i {\cal F}_E^{-1} {\cal F}_j = {\cal F}_j {\cal F}_E^{-1} {\cal F}_i  \qquad \qquad \forall i,j=1,\dots,n
\eeq
That ${\cal F}$ satisfies these equations can be proven directly using the associativity of $*_E$ together with the residue formula (\ref{residue Toda}).

There are other special K\"ahler manifolds, which are not the almost dual of a Frobenius manifold, for which this is also true. Among these is the periodic Toda chain, for which the equivalent of the metric $(.,.)$ is not flat. On the one hand such manifolds are weaker than almost Frobenius manifolds, on the other they carry more structure because they are also special K\"ahler. We approach this issue from the special K\"ahler side and introduce what we call extra special K\"ahler manifolds.
\begin{definition}
For any special K\"ahler manifold $B$ and an element $V\in \Gamma({\cal T}B)$ we define a holomorphic symmetric 2-tensor $(.,.)_V$ given by 
\eq
(X,Y)_V = \Xi(V,X,Y)
\eeq
\end{definition}

If ${\cal F}$ is homogeneous of degree 2, there is no choice of $V$ that makes the metric nondegenerate. We will exclude such cases and assume from now on that the metric is nondegenerate.
One defines the curvature of a holomorphic symmetric nondegenerate 2-tensor from its dependence of holomorphic coordinates, and as such the metric $(.,.)_V$ is in general not flat.

\begin{definition}
For any special K\"ahler manifold $B$ and an element $V\in \Gamma({\cal T}B)$ we define a commutative unital multiplication $*_V$ on each holomorphic tangent space ${\cal T}_bB$, varying smoothly with the point $b\in B$, by the following formula
\eq
\label{defmul}
( X, Y*_VZ )_V = \Xi (X,Y,Z)
\eeq
where the right hand side is independent of $V$.
Nondegeneracy of $(.,.)_V$ ensures that $*_V$ is well-defined, $*_V$ is unital with unit $V$, and $*_V$ and $( .,.)_V$ are compatible.
\end{definition}
The following lemma clarifies the role of the arbitrary $V$ somewhat.
\begin{lemma}
Different choices of $V$ correspond to rescalings of the algebra. In particular, whether or not the algebra is associative is independent of the choice. Nondegeneracy and flatness of $(.,.)_V$ may depend on it.
\end{lemma}
In the case of the open Toda chain, the choice $V=\sum_k z_k \partial_{z_k}$ leads to the intersection form of the corresponding simple singularity.

We are interested in associative multiplications. From the definition of a special K\"ahler manifold one finds that it is locally completely specified by any holomorphic prepotential ${\cal F}(z)$ whose second order derivatives give rise to a positive definite metric through the K\"ahler form (\ref{omega}). Associativity is the condition that ${\cal F}$ satisfy the WDVV equations, which is a highly nontrivial requirement that is certainly not met for an arbitrary integrable system. Nevertheless, many integrable systems have already been identified that do have an associative multiplication, see e.g. \cite{MARS-MIRO-MORO:2000, HOEV-MART2:2003, BRAD-MARS-MIRO-MORO:2006, HOEV:2007}.
\begin{definition}
Given a special K\"ahler manifold $(B,\omega,I,\nabla)$ and a choice of $V$ we define the metric $(.,.)_V$ and multiplication $*_V$ as above. If $(.,.)_V$ is nondegenerate and $*_V$ is associative we call the manifold {\bf{extra special K\"ahler}}. Note that this defines a Frobenius algebra structure on each of the holomorphic tangent spaces, varying smoothly with the point. These manifolds also have a potentiality property in the sense that there exist special coordinates $z_k$ and the special K\"ahler prepotential ${\cal F}(z)$ in terms of which
\eq
(\partial_{z_i},\partial_{z_j}*_V \partial_{z_k})_V = \frac{\partial^3{\cal F}}{\partial z_i \partial z_j \partial z_k}
\eeq
The prepotential satisfies the WDVV equations (\ref{gen WDVV}) with $E$ replaced by $V$.
\end{definition}
We will now make an attempt to place these manifolds amongst the various types of manifolds possessing multiplication on the tangent bundle, such as $F$-manifolds, $F$-manifolds with compatible flat structure, weak Frobenius manifolds, almost Frobenius manifolds and Frobenius manifolds. The original notion of course is that of Frobenius manifold, which has the most structure and was subsequently weakened for example by Manin and Hertling \cite{HERT-MANI:1999} to fit a more general context. It is not straightforward to place our manifold amongst them, mainly because the special K\"ahler connection $\nabla$ is {\it{real}} from the outset. All connections occurring in the range from $F$-manifolds to Frobenius manifolds are complex, which makes for a difficult comparison.
\begin{theorem}
If there exists a choice of $V$ such that $(.,.)_V$ is constant in the special coordinates, then the extra special K\"ahler manifold is an F-manifold with compatible flat metric. In Manin's terminology \cite{MANI:1999} this is a Frobenius manifold. In Dubrovin's terminology, what is missing is flatness of the identity and an Euler vector field. Since we are dealing with special K\"ahler manifolds that have this property, we will call them {\bf{special $F$-manifolds}}.
\end{theorem}
\begin{proof}
We check (\ref{id}) in the special coordinates, using $X,Y,Z,W=\partial_i,\partial_j,\partial_l,\partial_m$. The structure constants and ${\cal F}$ are related by
\eq
C_i = {\cal F}_i {\cal F}_V^{-1}  
\eeq
where $C_i$ is a matrix whose entries are the structure constants $C_{ij}^k$. Since $(.,.)_V$ is constant in the special coordinates, so is ${\cal F}_V$. Using $\partial_iC_j=\partial_jC_i$ and the WDVV equations $C_iC_j=C_jC_i$ repeatedly, one finds that the left hand side of (\ref{id}) amounts to
\begin{align}
\left[ C_i\partial_lC_m + (\partial_l C_i)C_m + (\partial_m C_i)C_l \right]_{jn}
-
\left[ C_l \partial_i C_j + (\partial_j C_l)C_i + (\partial_i C_l)C_j  \right]_{mn}
=\\
\partial_l \left[ C_iC_m \right]_{jn}
+
\left[  (\partial_iC_m)C_l \right]_{jn}
-
\partial_i \left[  C_lC_j\right]_{mn}
-
\left[ (\partial_l C_j)C_i \right]_{mn}
=\\
\partial_l \left[ C_iC_m-C_mC_i  \right]_{jn}
+
\partial_i \left[ C_jC_l-C_lC_j \right]_{mn}
+
\left[ C_j \partial_l C_i\right]_{mn}
-
\left[ C_m\partial_i C_l\right]_{jn}
=0
\end{align}
We now use theorem 2 in \cite{HERT-MANI:1999}, which states that $F$-manifolds with compatible flat metric are Frobenius manifolds in the sense of Manin \cite{MANI:1999}.
\end{proof}

\section{Conclusions and outlook}
We have given a short review on Frobenius manifolds and algebraic integrability, focusing on the overlap between these topics. From the Frobenius manifold point of view, almost duality assigns to the Frobenius manifold of a simple singularity an almost Frobenius manifold associated with the open Toda chain. In this paper we have defined a multiplication and compatible metric for any integrable system. Potentiality is ensured by the special K\"ahler structure but associativity of the multiplication and nondegeneracy and/or flatness of the metric are not. If we are dealing with a nondegenerate metric and an associative multiplication we we call this an extra special K\"ahler manifold. If moreover the metric is flat we call it a special $F$-manifold. This is a Frobenius manifold in the terminology of Manin \cite{MANI:1999} together with a special K\"ahler structure.

On the one hand one cannot expect an arbitrary Frobenius manifold to be special K\"ahler, since the requirement that the K\"ahler metric constructed from the prepotential be positive definite is nontrivial\footnote{We are concerned with global issues here. Since positive definiteness is an open condition it may be possible to define a germ of a special K\"ahler manifold.}.  
On the other hand we have argued that an arbitrary special K\"ahler manifold is no Frobenius manifold since the WDVV equations are nontrivial, and one may wonder how large the intersection is precisely.
The examples of extra special K\"ahler manifolds found thus far come from Seiberg-Witten theory and include various versions of Toda chains and spin chains. 
It is also known that the Neumann system, which is not related to Seiberg-Witten theory, is very special K\"ahler \cite{HOEV:2007}. Based on this observation, on the relation between systems of Toda \& Neumann type and Hitchin systems and on the residue formula found recently for Hitchin systems \cite{BALD:2006}, we conjecture that all Hitchin systems are very special K\"ahler. The most natural place to observe the $F$-manifold structure should be on the moduli space of certain (non-compact) Calabi-Yau varieties, which are related to Seiberg-Witten integrable systems on the one hand by geometric engineering (see e.g.  \cite{KLEM-LERC-MAYR-VAFA-WARN:1996, LERC-WARN:1998}), and to Hitchin systems on the other as suggested in the recent paper \cite{DIAC-DIJK-DONA-HOFM-PANT:2006}. The direct relation between Hitchin systems and Seiberg-Witten theory was observed in \cite{DONA-WITT:1996}.

\pagebreak

\bibliographystyle{amsalpha}
\bibliography{../biblio/biblio}
\include{thebibliography}

\end{document}

%% file: uppercycles.pdftex_t
\begin{picture}(0,0)%
\includegraphics{uppercycles}%
\end{picture}%
\setlength{\unitlength}{3947sp}%
\begingroup\makeatletter\ifx\SetFigFont\undefined%
\gdef\SetFigFont#1#2#3#4#5{%
  \reset@font\fontsize{#1}{#2pt}%
  \fontfamily{#3}\fontseries{#4}\fontshape{#5}%
  \selectfont}%
\fi\endgroup%
\begin{picture}(6698,3924)(160,-3316)
\put(901,-1074){\makebox(0,0)[lb]{\smash{{\SetFigFont{12}{14.4}{\familydefault}{\mddefault}{\updefault}{\color[rgb]{0,0,0}$\gamma_2$}%
}}}}
\put(1439,113){\makebox(0,0)[lb]{\smash{{\SetFigFont{12}{14.4}{\familydefault}{\mddefault}{\updefault}{\color[rgb]{0,0,0}$\gamma_1$}%
}}}}
\put(2700,-1550){\makebox(0,0)[lb]{\smash{{\SetFigFont{12}{14.4}{\familydefault}{\mddefault}{\updefault}{\color[rgb]{0,0,0}$\gamma_3$}%
}}}}
\put(4377,-1361){\makebox(0,0)[lb]{\smash{{\SetFigFont{12}{14.4}{\familydefault}{\mddefault}{\updefault}{\color[rgb]{0,0,0}$\gamma_4$}%
}}}}
\put(2990,339){\makebox(0,0)[lb]{\smash{{\SetFigFont{12}{14.4}{\familydefault}{\mddefault}{\updefault}{\color[rgb]{0,0,0}$\delta_1$}%
}}}}
\put(2976,-336){\makebox(0,0)[lb]{\smash{{\SetFigFont{12}{14.4}{\familydefault}{\mddefault}{\updefault}{\color[rgb]{0,0,0}$\delta_2$}%
}}}}
\put(3825,-823){\makebox(0,0)[lb]{\smash{{\SetFigFont{12}{14.4}{\familydefault}{\mddefault}{\updefault}{\color[rgb]{0,0,0}$\delta_3$}%
}}}}
\put(5714,-724){\makebox(0,0)[lb]{\smash{{\SetFigFont{12}{14.4}{\familydefault}{\mddefault}{\updefault}{\color[rgb]{0,0,0}$\delta_4$}%
}}}}
\put(440,-2746){\makebox(0,0)[lb]{\smash{{\SetFigFont{12}{14.4}{\rmdefault}{\mddefault}{\updefault}{\color[rgb]{0,0,0}$\zeta_+(x)$}%
}}}}
\end{picture}%

%% file: cycles.pdftex_t
\begin{picture}(0,0)%
\includegraphics{cycles}%
\end{picture}%
\setlength{\unitlength}{3947sp}%
\begingroup\makeatletter\ifx\SetFigFont\undefined%
\gdef\SetFigFont#1#2#3#4#5{%
  \reset@font\fontsize{#1}{#2pt}%
  \fontfamily{#3}\fontseries{#4}\fontshape{#5}%
  \selectfont}%
\fi\endgroup%
\begin{picture}(6699,3924)(1714,-3748)
\put(3376,-961){\makebox(0,0)[lb]{\smash{{\SetFigFont{12}{14.4}{\familydefault}{\mddefault}{\updefault}{\color[rgb]{0,0,0}$x_1$}%
}}}}
\put(2776,-2086){\makebox(0,0)[lb]{\smash{{\SetFigFont{12}{14.4}{\familydefault}{\mddefault}{\updefault}{\color[rgb]{0,0,0}$x_2$}%
}}}}
\put(4651,-2536){\makebox(0,0)[lb]{\smash{{\SetFigFont{12}{14.4}{\familydefault}{\mddefault}{\updefault}{\color[rgb]{0,0,0}$x_3$}%
}}}}
\put(6301,-2311){\makebox(0,0)[lb]{\smash{{\SetFigFont{12}{14.4}{\familydefault}{\mddefault}{\updefault}{\color[rgb]{0,0,0}$x_4$}%
}}}}
\put(4576,-961){\makebox(0,0)[lb]{\smash{{\SetFigFont{12}{14.4}{\familydefault}{\mddefault}{\updefault}{\color[rgb]{0,0,0}$\delta_2$}%
}}}}
\put(5626,-1411){\makebox(0,0)[lb]{\smash{{\SetFigFont{12}{14.4}{\familydefault}{\mddefault}{\updefault}{\color[rgb]{0,0,0}$\delta_3$}%
}}}}
\put(7351,-1336){\makebox(0,0)[lb]{\smash{{\SetFigFont{12}{14.4}{\familydefault}{\mddefault}{\updefault}{\color[rgb]{0,0,0}$\delta_4$}%
}}}}
\put(5101,-61){\makebox(0,0)[lb]{\smash{{\SetFigFont{12}{14.4}{\familydefault}{\mddefault}{\updefault}{\color[rgb]{0,0,0}$\delta_1$}%
}}}}
\put(4726,-2911){\makebox(0,0)[lb]{\smash{{\SetFigFont{12}{14.4}{\familydefault}{\mddefault}{\updefault}{\color[rgb]{0,0,0}$\gamma_3$}%
}}}}
\put(2776,-2461){\makebox(0,0)[lb]{\smash{{\SetFigFont{12}{14.4}{\familydefault}{\mddefault}{\updefault}{\color[rgb]{0,0,0}$\gamma_2$}%
}}}}
\put(6376,-2686){\makebox(0,0)[lb]{\smash{{\SetFigFont{12}{14.4}{\familydefault}{\mddefault}{\updefault}{\color[rgb]{0,0,0}$\gamma_4$}%
}}}}
\put(3226,-211){\makebox(0,0)[lb]{\smash{{\SetFigFont{12}{14.4}{\familydefault}{\mddefault}{\updefault}{\color[rgb]{0,0,0}$\gamma_1$}%
}}}}
\put(7051,-361){\makebox(0,0)[lb]{\smash{{\SetFigFont{12}{14.4}{\familydefault}{\mddefault}{\updefault}{\color[rgb]{0,0,0}$x_{5}$}%
}}}}
\end{picture}%

%% file: lowercycles.pdftex_t
\begin{picture}(0,0)%
\includegraphics{lowercycles}%
\end{picture}%
\setlength{\unitlength}{3947sp}%
\begingroup\makeatletter\ifx\SetFigFont\undefined%
\gdef\SetFigFont#1#2#3#4#5{%
  \reset@font\fontsize{#1}{#2pt}%
  \fontfamily{#3}\fontseries{#4}\fontshape{#5}%
  \selectfont}%
\fi\endgroup%
\begin{picture}(6701,3923)(161,-7430)
\put(470,-6844){\makebox(0,0)[lb]{\smash{{\SetFigFont{12}{14.4}{\rmdefault}{\mddefault}{\updefault}{\color[rgb]{0,0,0}$\zeta_-(x)$}%
}}}}
\end{picture}%